\documentclass[usenatbib]{mn2e}
\usepackage{natbibmnfix,graphicx,times}

\newcommand{\colden}{\mbox{ cm$^{-2}$}}
\newcommand{\Mpc}{\mbox{ Mpc}}

\newcommand{\kpc}{\mbox{ kpc}}

\newcommand{\sqm}{\mbox{ m$^2$}}

\newcommand{\keV}{\mbox{ keV}}

\newcommand{\kel}{\mbox{ K}}
\newcommand{\mkel}{\mbox{ mK}}

\newcommand{\mJy}{\mbox{ mJy}}

\newcommand{\MHz}{\mbox{ MHz}}
\newcommand{\kHz}{\mbox{ kHz}}

\newcommand{\hunits}{\mbox{ km s$^{-1}$ Mpc$^{-1}$}}

\newcommand{\Junits}{\mbox{ cm$^{-2}$ s$^{-1}$ Hz$^{-1}$ sr$^{-1}$}}

\newcommand{\bxhi}{\bar{x}_{\rm HI}}
\newcommand{\xhi}{x_{\rm HI}}

\newcommand{\bxion}{\bar{x}_i}

\newcommand{\dtb}{\delta T_b}

\newcommand{\htwo}{HII }
\newcommand{\mmin}{m_{\rm min}}
\newcommand{\fcoll}{f_{\rm coll}}
\newcommand{\fesc}{f_{\rm esc}}
\newcommand{\lya}{Ly$\alpha$ }
\newcommand{\deriv}{{\rm d}}
\newcommand{\apj}{ApJ \ }
\newcommand{\apjl}{ApJ \ }

\newcommand{\aap}{A\&A \ }
\newcommand{\aj}{AJ \ }
\newcommand{\mnras}{MNRAS \ }
\newcommand{\pasj}{PASJ  \ }
\newcommand{\physrep}{Physics Reports \ }

\title[The 21 cm forest]{The 21 centimeter forest}

\author[S.~R. Furlanetto]{Steven R.  Furlanetto\thanks{Email: steven.furlanetto@yale.edu} \\
Yale Center for Astronomy and Astrophysics, Yale University, 260 Whitney Avenue, New Haven, CT 06520-8121}

\begin{document}

\maketitle

\begin{abstract}
We examine the prospects for studying the pre-reionization intergalactic medium (IGM) through the so-called 21 cm forest in spectra of bright high-redshift radio sources.  We first compute the evolution of the mean optical depth $\tau$ for models that include X-ray heating of the IGM gas, Wouthuysen-Field coupling, and reionization.  Under most circumstances, the spin temperature $T_S$ grows large well before reionization begins in earnest.  As a result, $\tau \la 10^{-3}$ throughout most of reionization, and background sources must sit well beyond the reionization surface in order to experience measurable absorption.  \htwo regions produce relatively large ``transmission gaps" and may therefore still be observable during the early stages of reionization.  Absorption from sheets and filaments in the cosmic web fades once $T_S$ becomes large and should be rare during reionization.  Minihalos can produce strong (albeit narrow) absorption features.  Measuring their abundance would yield useful limits on the strength of feedback processes in the IGM as well as their effect on reionization.
\end{abstract}


\section{Introduction} \label{intro}

The ``twilight zone" of structure formation, when the first galaxies and quasars formed and eventually reionized the Universe, is the last frontier of galaxy formation studies.  Observations are now beginning to probe this epoch, through both direct observations of high-redshift galaxies (e.g., \citealt{bouwens04, taniguchi05}) and indirect constraints on reionization (e.g., \citealt{fan01, malhotra04, fan06, page06}).  But the chief lesson of these initial forays has been how little we know about  the first luminous sources, their formation processes, their feedback on the intergalactic medium (IGM), and reionization itself (see, e.g., \citealt{barkana01, ciardi05-rev}).

Thus a great deal of attention is now focused on developing new approaches to observing the $z \ga 10$ Universe.  Perhaps the most exciting is the 21 cm transition of neutral hydrogen (see \citealt{furl06-review} and references therein).  Of the many applications of this transition, the most popular possibility is to perform ``21 cm tomography" of the IGM by mapping the redshift-space fluctuations in its brightness temperature \citep{hogan79, scott90, madau97, zald04}.  By measuring the entire history of reionization, as well as its spatial structure, such observations could provide extraordinarily powerful constraints on reionization and early structure formation.  Unfortunately, this technique also suffers from three intrinsic difficulties.  First, it requires the spin temperature of the IGM, $T_S$, to differ from the cosmic microwave background (CMB) temperature $T_\gamma$.  Otherwise, the two blend together and cannot be differentiated.  Unfortunately, this condition is not necessarily satisfied in the earliest phases of structure formation.  Second, given realistic telescope parameters, tomography cannot resolve structures smaller than $\sim 1 \Mpc$, which is well above the size of collapsed (or collapsing) structures at $z>10$.  It therefore does not probe structure formation across all scales.  Finally, tomography presents a number of truly formidable observational challenges, such as foreground contamination, ionospheric distortions, and terrestrial interference.

In this paper, we will consider a complementary probe immune to most of these problems:  the ``21 cm forest" \citep{carilli02, furl02-21cm}.  The technique is the exact analog of the \lya forest that is so useful for studying the IGM at $z \la 6$.  Neutral hydrogen along the line of sight to a distant radio-loud quasar resonantly absorbs continuum radiation that redshifts into the local 21 cm transition.  Thus a high resolution, high signal-to-noise spectrum can map the fluctuating neutral hydrogen field along this line of sight.  Interestingly, the 21 cm forest has a surprisingly long history:  \citet{field59-obs} attempted to observe it toward Cygnus A several years before the \lya forest was even discovered (in the process, he ruled out an $\Omega=1$ universe full of neutral hydrogen).

The 21 cm forest bypasses all of the difficulties with tomographic measurements.   First, the CMB no longer acts as the background source, so the IGM remains visible even if $T_S = T_\gamma$.  Second, the IGM can be resolved spatially on fine scales (tens of kpc) given a sufficiently bright background source, allowing us to probe individual filaments and minihalos with square-kilometer class telescopes.  Finally, spectroscopy of bright point sources is relatively straightforward from an observer's perspective, and 21 cm forest observations are immune to many of the systematics that beset tomography.

We expect to find four types of absorption features.  First, the mean level of absorption measures the global evolution of the radiation background \citep{carilli02}.  Second, sheets and filaments in the ``proto-cosmic web" produce stronger absorption features, just as they do in the \lya forest:  these features provide insight into the first phases of structure formation \citep{carilli02}.  Third, a fair fraction of the IGM gas may condense into ``minihalos" before reionization.  These are virialized neutral clumps that are too small to fragment and form stars but may have important implications for reionization \citep{oh03-entropy, iliev05, ciardi05-mh}.  No other method has been devised to detect them directly, but they produce relatively strong 21 cm absorption \citep{furl02-21cm}.  Finally, \htwo regions produce transmission gaps in the absorption spectrum; their evolution would obviously constrain the evolution of the ionizing background as well as its spatial structure. While the first three of these classes have been considered before, they have not been examined in the context of self-consistent histories of the 21 cm background.  The purpose of this paper is to re-examine each kind of absorber in light of this broader context.  We will examine the features each imprints on the 21 cm forest, the epochs at which they are important, and their observable implications.

The remainder of this paper is organized as follows.  We describe some basic properties of the 21 cm transition in \S \ref{21cm}.  We then discuss each of the four types of spectral features in turn:  the mean level in \S \ref{global}, \htwo regions in \S \ref{htwo}, the cosmic web in \S \ref{web}, and minihalos in \S \ref{minihalo}.  We conclude and discuss potential observations in \S \ref{disc}.

In our numerical calculations, we assume a cosmology with $\Omega_m=0.26$, $\Omega_\Lambda=0.74$, $\Omega_b=0.044$, $H=100 h \hunits$ (with $h=0.74$), $n=0.95$, and $\sigma_8=0.8$, consistent with the most recent measurements \citep{spergel06}, although we have increased $\sigma_8$ from the best-fit \emph{WMAP} value to improve agreement with weak lensing.  We quote all distances in comoving units, unless otherwise specified.

\section{The 21 cm Transition} \label{21cm}

We review the relevant characteristics of the 21 cm transition here; we refer the interested reader to \citet{furl06-review} for a more comprehensive discussion.  The 21 cm optical depth of a patch of the IGM is \citep{field59-obs}
\begin{eqnarray}
\tau & \approx  & 9.6 \times 10^{-3} \, \xhi \, (1+\delta) \, \left( \frac{1+z}{10} \right)^{3/2} \, \left[ \frac{T_\gamma(z)}{T_S} \right] \nonumber \\
& & \times \left[ \frac{H(z)/(1+z)}{\deriv v_\parallel/\deriv r_\parallel} \right],
\label{eq:tauforest}
\end{eqnarray}
where $\xhi$ is the neutral fraction, $\delta$ is the fractional overdensity, $T_\gamma$ is the CMB temperature at redshift $z$, $T_S$ is the spin temperature, and $\deriv v_\parallel/\deriv r_\parallel$ is the line of sight velocity in physical units.  The brightness temperature of the patch viewed against a source of brightness temperature $T_{\rm rg}$ is (neglecting peculiar velocities for simplicity)
\begin{eqnarray}
\dtb & = & 27 \xhi \, (1 + \delta) \, \left( \frac{\Omega_b h^2}{0.023} \right) \left( \frac{0.15}{\Omega_m h^2} \, \frac{1+z}{10} \right)^{1/2} \nonumber \\
& & \times \left( \frac{T_S - T_{\rm rg}}{T_S} \right) \mkel.
\label{eq:dtb}
\end{eqnarray}
Note that $\dtb \propto T_{\rm rg}/T_S$ for $T_{\rm rg} \gg T_S$, the regime relevant for this paper.  Thus the IGM temperature is crucial to estimating the signal strength.

Equation~(\ref{eq:tauforest}) applies only to the expanding IGM, because it uses the expansion rate to determine the total column.  For a single isolated cloud, the central optical depth is 
\begin{equation}
\tau_{\rm cl} \approx 0.076 \, \left( \frac{10^3 \kel}{T_K} \right)^{1/2} \left( \frac{10^3 \kel}{T_S} \, \frac{N_{\rm HI}}{10^{21} \colden} \right),
\label{eq:taucloud}
\end{equation}
where $T_K$ is the kinetic temperature, $N_{\rm HI}$ is the total column density of neutral hydrogen, and we have assumed pure thermal broadening.  Thus, clouds must have column densities close to those of damped-\lya absorbers in order to cause substantial 21 cm absorption.

The spin temperature $T_S$ is determined by a competition between three processes:  atomic collisions, scattering of CMB photons, and scattering of Ly$\alpha$ photons \citep{wouthuysen52, field58}.  In equilibrium,
\begin{equation}
T_S^{-1} = \frac{T_\gamma^{-1} + \tilde{x}_\alpha \tilde{T}_c^{-1} + x_c T_K^{-1}}{1 + \tilde{x}_\alpha + x_c}.
\label{eq:tsdefn}
\end{equation}
Here $x_c$ is the sum of the collisional coupling coefficients for H--H interactions \citep{zygelman05} and H-e$^-$ collisions \citep{smith66, liszt01}; it is of course proportional to the local density.   The middle term describes the Wouthuysen-Field effect, in which absorption and re-emission of Ly$\alpha$ photons mix the hyperfine states.  The coupling coefficient is
\begin{equation}
\tilde{x}_\alpha = 1.81 \times 10^{11} (1+z)^{-1} \tilde{S}_\alpha J_\alpha,
\label{eq:xalpha}
\end{equation}
where $\tilde{S}_\alpha$ is a factor of order unity describing the detailed atomic physics of the scattering process \citep{chen04, hirata05} and $J_\alpha$ is the background flux at the Ly$\alpha$ frequency in units $\Junits$; the Wouthuysen-Field effect becomes efficient when there are $\sim 0.1$--$1$ photons per baryon near this frequency.  It couples $T_S$ to an effective color temperature $\tilde{T}_c$; typically $\tilde{T}_c \approx T_K$ \citep{field59-ts}.  We use the numerical fits of \citet{hirata05} for $\tilde{S}_\alpha$ and $\tilde{T}_c$.

\section{The Average Signal} \label{global}

We will first consider the evolution of the mean optical depth with time, which is determined purely by the mean neutral fraction $\bxhi(z)$ and $T_S(z)$; note that we neglect fluctuations in these quantities here.  These depend on the thermal history of the IGM (including adiabatic expansion, X-ray heating, and shocks), the ultraviolet (UV) background (which determines the \lya coupling and hence spin temperature), and the ionizing efficiency.  Unfortunately, none of these factors are well-constrained at high redshifts.  Rather than attempt to make robust predictions, we will therefore content ourselves with examining a range of simple histories to develop some intuition about the qualitative features.  

We will use the simple model of \citet{furl06-glob} to compute the optical depth history; we refer the reader there for a detailed discussion of the various prescriptions.  In brief, this model assumes that stars dominate the radiation background and sets the total star formation rate via the rate at which gas collapses onto galaxies, $\deriv \fcoll/\deriv t$, where $\fcoll$ is the collapse fraction (or the mass fraction in halos with virial temperatures $T_{\rm vir} > 10^4 \kel$).  The ionizing efficiency is $\zeta = A_{\rm He} f_\star f_{\rm esc} N_{\rm ion}$, where $f_\star$ is the star formation efficiency, $f_{\rm esc}$ is the escape fraction of ionizing photons, $N_{\rm ion}$ is the number of ionizing photons produced per baryon incorporated into stars, and $A_{\rm He}$ is a normalization constant accounting for helium in the IGM.  

%%%%%%%%%%%% FIGURE 1: Mean optical depth, single population
\begin{figure*}[!t]
\begin{center}
\resizebox{8cm}{!}{\includegraphics{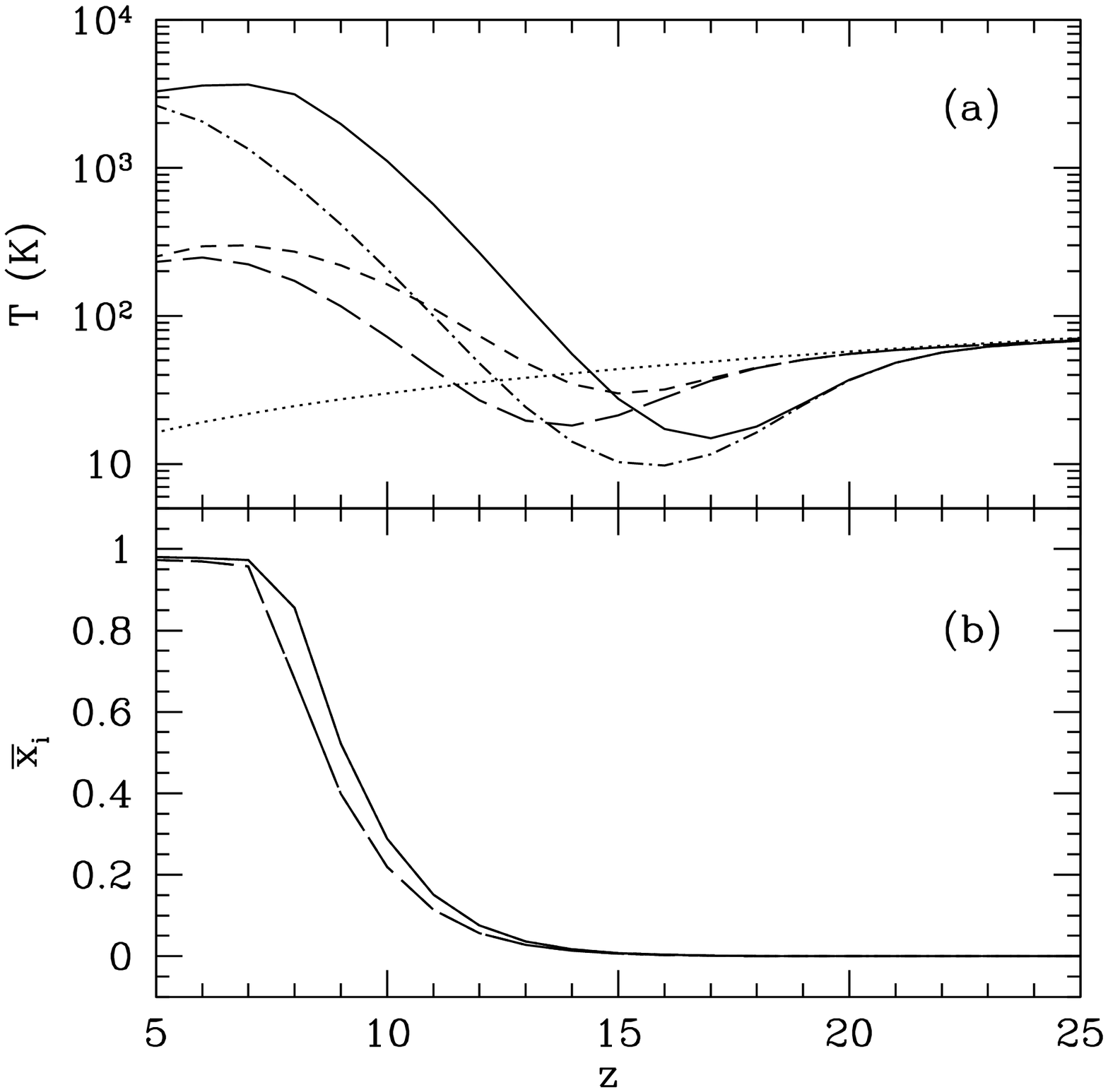}}
\hspace{0.13cm}
\resizebox{8cm}{!}{\includegraphics{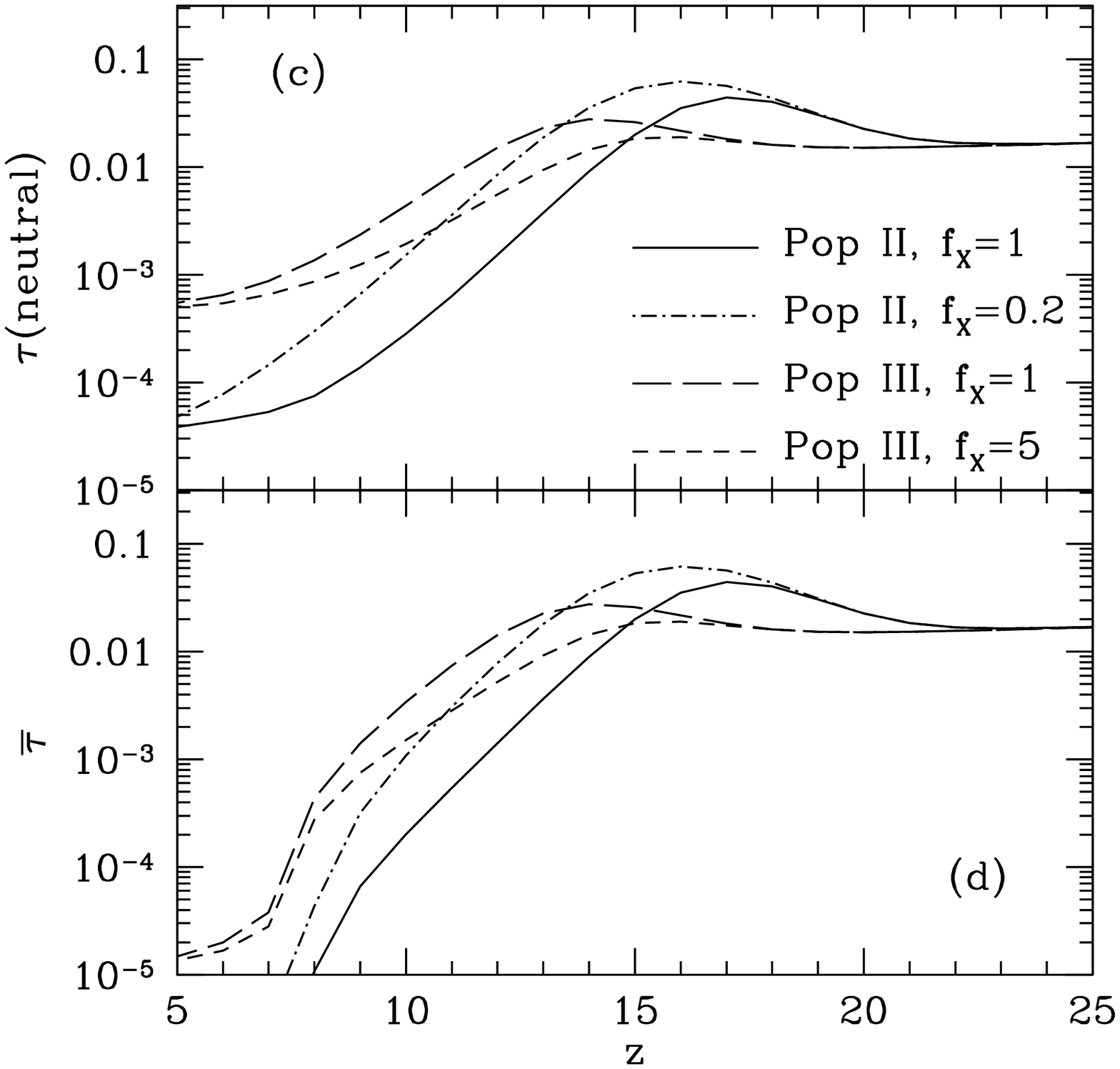}}
\end{center}
%\plottwo{f1ab.eps}{f1cd.eps}
\caption{Optical depth evolution in one-population reionization histories.  \emph{(a)}:  Spin temperature.   The dotted curve shows $T_\gamma$ for reference. \emph{(b)}: Ionized fraction  \emph{(c)}: Optical depth in neutral regions \emph{(d)}: Mean optical depth.  The solid and dot-dashed curves use our fiducial Pop II model with $f_X=1$ and $0.2$, respectively.  The long- and short-dashed curves use the Pop III model with $f_X=1$ and $5$, respectively. }
\label{fig:taumean}
\end{figure*}

The rate of X-ray heating is also assumed to be proportional to $\deriv \fcoll/\deriv t$, with a fiducial proportionality constant fixed by the local relation between star formation rate and X-ray luminosity \citep{grimm03,ranalli03,gilfanov04}.  We let $f_X$ measure the efficiency at high redshifts relative to this local value (measured between $0.2$ and $10 \keV$).  We include a simple model for shock heating from \citet{furl04-sh} that describes shocks in the high-redshift cosmic web reasonably well \citep{kuhlen06-21cm}.

The final element is the UV background responsible for the Wouthuysen-Field effect.  We again assume that the emissivity is proportional to $\deriv \fcoll/\deriv t$ and use the (low-metallicity) Population II and (very massive) Population III (henceforth Pop II and Pop III) spectral fits of \citet{barkana05-ts} to estimate the efficiency with which these photons are produced.  Note that we properly incorporate higher Lyman-series transitions \citep{hirata05, pritchard05}.  

Although there are obviously a number of free parameters in this model, the general features are easy to understand.  The 21 cm background contains three major transitions:  the points when $T_K$ first exceeds $T_\gamma$, when \lya coupling becomes efficient, and when reionization occurs.  Because it affects all the backgrounds equally, $f_\star$ simply shifts everything forward and backward in time.  The X-ray efficiency $f_X$ only affects the kinetic temperature, while $f_{\rm esc}$ only affects $\bxhi$.  The stellar initial mass function -- especially the choice between Pop II and very massive Pop III stellar populations -- affects the ratio of \lya photons to ionizing photons and moves the onset of \lya coupling relative to reionization.  Because they are so much hotter, Pop III stars push this transition forward in time, compressing the features in the 21 cm background. 

Figure~\ref{fig:taumean} shows some results for single-population models (without feedback).  For Pop II stars, we take $f_\star=0.1$, $f_{\rm esc}=0.1$, $N_{\rm ion}=4000$, and $f_X=1$ (solid curves).  For Pop III stars, we take $f_\star=0.01$, $\fesc=0.1$, $f_X=1$, $N_{\rm ion}=30,000$, and $f_X=1$ (long-dashed curves), for a somewhat smaller overall ionizing efficiency.  We also show results for Pop II stars with a low X-ray heating efficiency $f_X=0.2$ (dot-dashed curves) and Pop III stars with a high X-ray efficiency $f_X=5$ (short-dashed curves).  

Panels \emph{a} and \emph{b} show $T_S(z)$ and $\bxion(z) = 1 - \bxhi(z)$.  With this normalization, reionization ends at $z \sim 7$.  The most important point is that $T_S$ increases beyond $T_\gamma$ well before reionization completes, although the transition does occur significantly later for Pop III stars.  Crucially, $T_S$ also saturates at a smaller value for these stars.  

%%%%%%%%%%%% FIGURE 2: Mean optical depth, feedback
\begin{figure*}[!t]
\begin{center}
\resizebox{8cm}{!}{\includegraphics{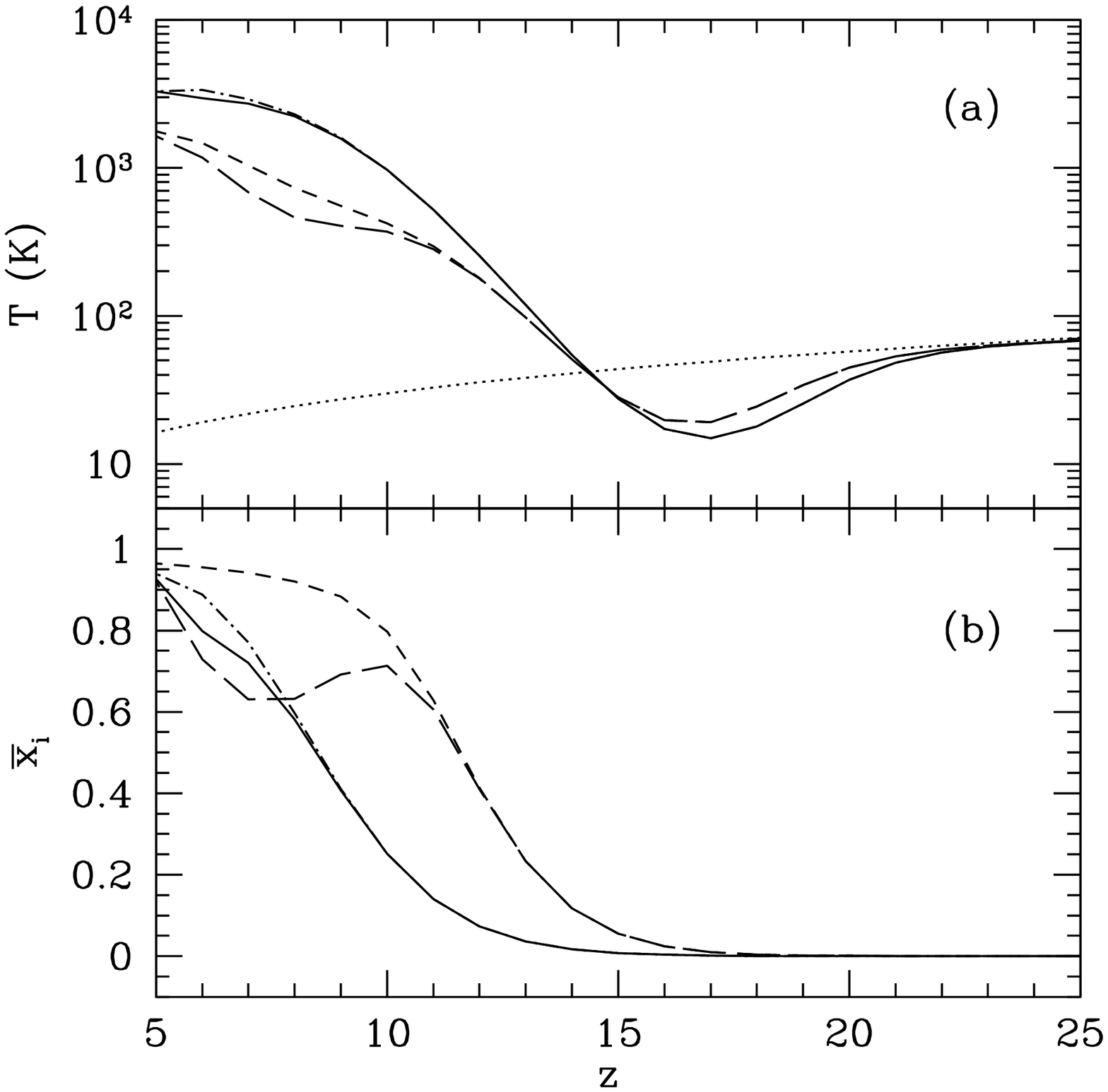}}
\hspace{0.13cm}
\resizebox{8cm}{!}{\includegraphics{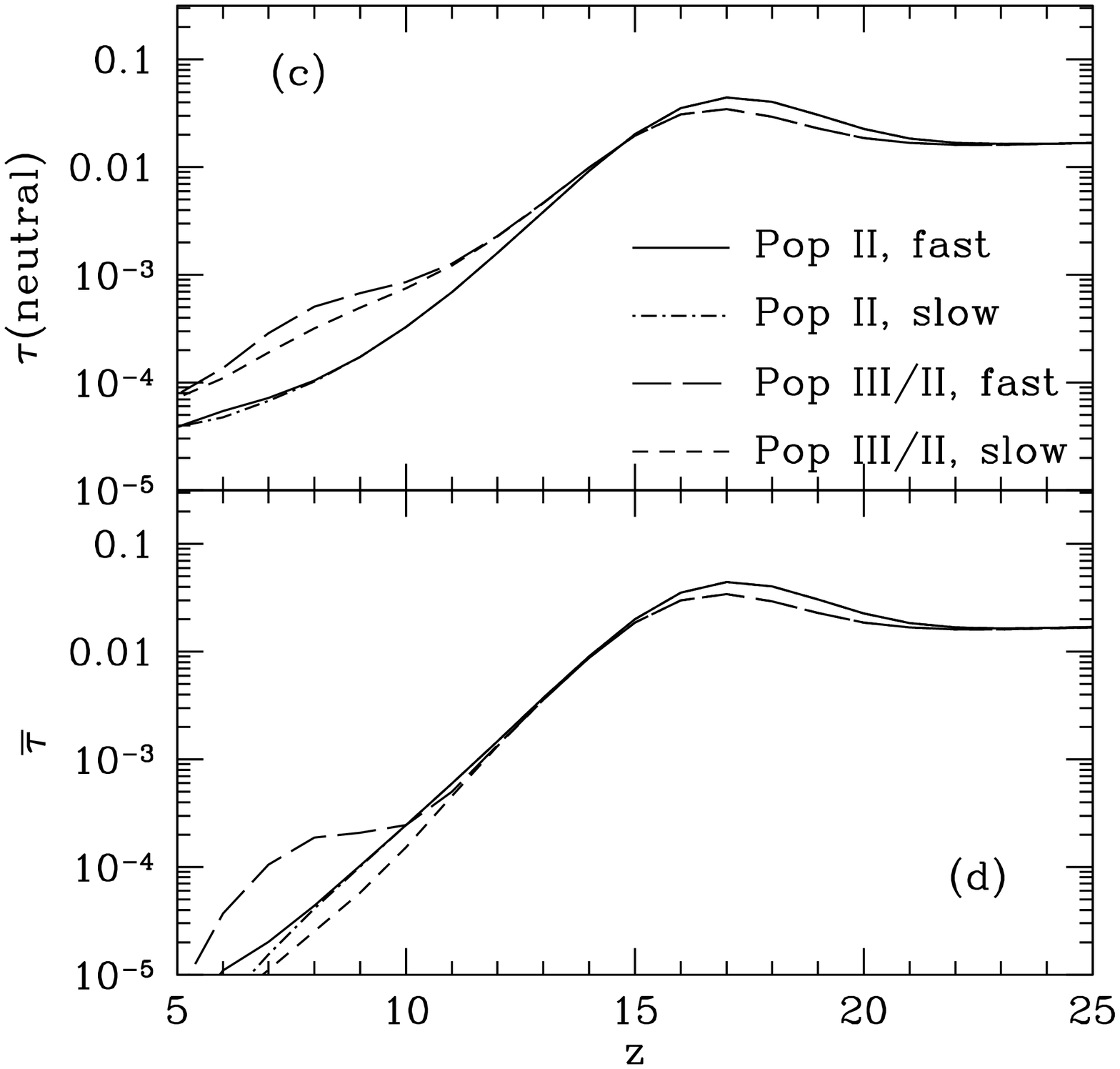}}
\end{center}
%\plottwo{f2ab.eps}{f2cd.eps}
\caption{As Figure~\ref{fig:taumean}, but for models that include photoheating feedback.  The solid and dot-dashed curves use our fiducial Pop II parameters.  The long and short-dashed curves assume that Pop III (Pop II) stars form in cold (hot) regions.    In each case, the two curves take different feedback speeds (see text).  Note that the ``double reionization" seen in the long-dashed curve is difficult to achieve in realistic situations.}
\label{fig:taumean-fb}
\end{figure*}

Panel \emph{c} shows the average optical depth in neutral regions (i.e., the contrast between neutral absorbing regions and empty \htwo regions).  Panel \emph{d} shows the mean optical depth as a function of redshift (i.e., the optical depth averaged over long stretches of the spectrum, including both neutral and ionized regions).  At sufficiently high redshifts, $T_S \approx T_\gamma$ and $\bar{\tau}$ evolves slowly.  Once \lya coupling becomes strong, $\bar{\tau}$ increases somewhat because (in all of these models) X-ray heating kicks in somewhat later.  However,  it does not take long for X-ray heating to dominate the temperature.  From this point, the optical depth decreases rapidly:  for our fiducial Pop II parameters, $\bar{\tau} \la 10^{-3}$ almost before reionization begins.  Even though massive Pop III stars produce comparable kinetic temperatures, the IGM optical depth remains about an order magnitude larger if they dominate.  This is purely a function of the spin temperature (and lack of \lya photons), so it would also occur if quasars dominated the radiation background.  Thus the 21 cm forest may be a more powerful probe if Pop III stars, or some other hard source, dominate.  Note that emission against the CMB does not depend on $T_S$ so long as $T_S \gg T_\gamma$.  Thus comparing the mean levels of emission and absorption could help to identify the type of sources during reionization.

In any case, Figure~\ref{fig:taumean} illustrates a crucial point about the 21 cm forest:  observing high optical depth features requires sources that shine well before reionization.  At $z \ga 15$, $\bar{\tau}$ can indeed be quite large -- nearly $10\%$ in some models -- but during reionization it is likely to be up to three orders of magnitude smaller. Unfortunately, this presents a quandary:  radio sources are likely to be associated with ionizing sources, so they too will be extremely rare at such high redshifts.  The 21 cm forest has the inherent difficult that the strongest absorption can only occur for the most distant sources, while it may be nearly unobservable for the most common sources during reionization.  

Of course, feedback (among other processes) could easily cause more complicated histories.  Figure~\ref{fig:taumean-fb} shows some examples.   The solid and dot-dashed curves assume our Pop II parameters but include strong photoheating feedback\footnote{Specifically, we assume that photoheating increases the minimum galaxy mass to $T_{\rm vir}=2 \times 10^5 \kel$ \citep{thoul96, kitayama00}, even though recent work suggests photoheating is actually less efficient \citep{dijkstra04-feed}.} (with varying speeds; see \citealt{furl05-double} and \citealt{furl06-glob} for details).  The long and short-dashed curves increase the effects of feedback by assuming that photoheating is accompanied by a simultaneous change in stellar population; again we allow two different feedback speeds.  The parameters for each population take our fiducial values, except that we set $f_\star=0.1$ for Pop III stars in order to exaggerate the decline in ionizing efficiency.  Note that the long-dashed curve yields a short period in which recombinations dominate.  This is unlikely to occur in reality \citep{furl05-double}, as we have stretched several parameters to achieve it (raised the Jeans mass in photoheated region to its maximal value, assumed unrealistically fast feedback, and conflated the transition between stellar populations with photoheating).  But it helps to illustrate how such an era might affect the signal.

The principal effect of feedback is to prolong reionization by $\Delta z \sim 5$--$10$.  One might then hope for a long period of strong contrast between neutral and ionized regions.  But panel \emph{c} shows that the optical depth in neutral gas nevertheless falls to $\sim 10^{-3}$ by $z \approx 10$ in all cases.  This is because X-ray heating, and \lya coupling, establish themselves in the early stages of reionization, when feedback is irrelevant.  Thus for a fixed endpoint to reionization, feedback weakens the 21 cm forest by pushing the X-ray and \lya coupling transitions farther back in time.  Again, the prospects are somewhat better if Pop III stars dominate the early stages.  

Now that we have some idea of the optical depths involved, it is useful to estimate their observability.  The minimum background source flux density required to detect an absorption feature with signal to noise S/N is
\begin{eqnarray}
S_{\rm min} & = & 16 \mJy \, \left( \frac{{\rm S/N}}{5} \, \frac{10^{-3}}{\tau} \, \frac{10^6 \sqm}{A_{\rm eff}} \, \frac{T_{\rm sys}}{400 \kel} \right) \nonumber \\
& & \times \left( \frac{100 \kHz}{\Delta \nu_{\rm ch}} \, \frac{1 {\rm \ week}}{t_{\rm int}} \right)^{1/2},
\label{eq:forest-sens}
\end{eqnarray}
where $A_{\rm eff}$ is the effective area of the telescope, $T_{\rm sys}$ is the system temperature, $\Delta \nu_{\rm ch}$ is the bandwidth of each channel (assumed to be narrower than the feature of interest), and $t_{\rm int}$ is the total ``on-source" integration time.  Here we have assumed simultaneous observation of two orthogonal polarizations, which is normal for dipole antennae, and taken telescope parameters similar to those expected for the SKA observing at $z=10$ (see also \citealt{carilli02}).  At these low frequencies, $T_{\rm sys}$ is dominated by the Galactic synchrotron background; on a quiet portion of the sky, this is typically (e.g., \citealt{furl06-review})
\begin{equation}
T_{\rm sky} \sim 180 \ \left( \frac{\nu}{180 \MHz} \right)^{-2.6} \kel.
\label{eq:tsky}
\end{equation}
As we shall see, observing any of the absorbers we will discuss requires rather luminous sources.  For example, the $z=10$ equivalent of Cygnus A would have $S \la 20 \mJy$ \citep{carilli02}.  

\section{\htwo regions} \label{htwo}

Ionized bubbles create gaps in the absorption.  In this section, we will estimate the abundance and sizes of such transmission gaps throughout reionization.  \citet{carilli02} implicitly included \htwo regions in their simulation of the 21 cm forest.  However, their box was only $4 h^{-1} \Mpc$ across.  Recent analytic models \citep{furl04-bub} and simulations \citep{iliev05-sim, zahn06-comp} have shown that ionized bubbles rapidly exceed this size, so much larger boxes -- or analytic models -- are required to sample the bubble distribution reliably.

%%%%%%%%%%%% FIGURE 3: Ionized regions
\begin{figure}
\begin{center}
\resizebox{8cm}{!}{\includegraphics{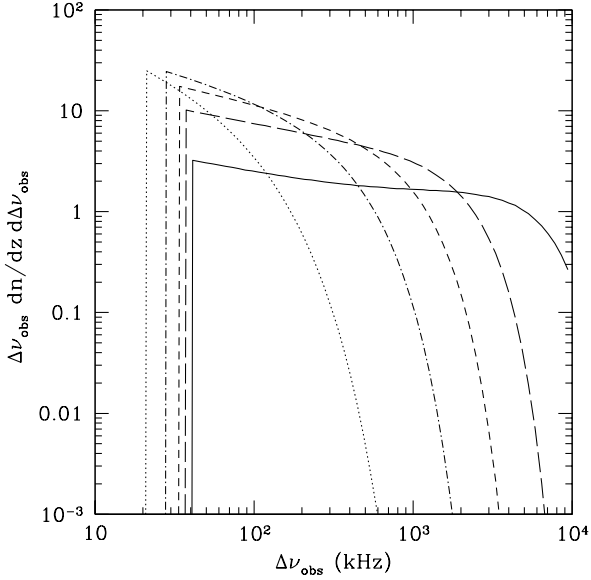}}
\end{center}
%\plotone{f3.eps}
\caption{Differential abundance of \htwo regions intersected per unit redshift and per unit radial width at $z=10$.  The dotted, dot-dashed, short-dashed, long-dashed, and solid lines take $\bxion=0.1,\,0.3,\,0.5,\,0.7$, and $0.9$, respectively.  }
\label{fig:htwo}
\end{figure}

We will therefore use a simple analytic model for the bubble size distribution \citep{furl04-bub, furl05-charsize} that provides a good match to simulations \citep{zahn06-comp}.  The model assumes that the distribution of ionized material is driven primarily by the strong clustering of ionizing sources.  We will ignore recombinations (or more precisely their spatial fluctuations, which appears to be a good assumption during the early and middle stages of reionization for which the 21 cm absorption signal is large; \citealt{furl05-rec}).  In that case, the mass that any galaxy can ionize is simply $\zeta m_{\rm gal}$.  Thus, an isolated region is fully ionized if $\fcoll > 1/\zeta$; because the collapse fraction is itself a function of the large-scale density field, this condition can be used to specify the ionization field.  \citet{furl04-bub} showed that the comoving number density of ionized bubbles $n_b(m)$ is then 
\begin{equation}
m \, n_b(m) = \sqrt{\frac{2}{\pi}} \, \frac{\bar{\rho}}{m} \, \left| \frac{ \deriv \ln \sigma}{\deriv \ln m} \right| \, \frac{B_0}{\sigma(m)} \, \exp \left[ - \frac{ B^2(m,z) }{ 2 \sigma^2(m) } \right],
\label{eq:nbub}
\end{equation}
where $B_0$ and $B$ are density thresholds from the condition on the collapse fraction, $\sigma^2(m)$ is the variance of the density field smoothed on mass $m$, and $\bar{\rho}$ is the mean comoving matter density.  Equation (\ref{eq:nbub}) can easily be generalized to allow the ionizing efficiency to vary with galaxy mass \citep{furl05-charsize}, so that $\zeta=\zeta(m_h)$ and $m_h$ is the halo mass.  As we increase the importance of massive galaxies, the bubbles grow larger, especially during the early phases of reionization.

With equation (\ref{eq:nbub}), the number density of transmission gaps in a 21 cm forest spectrum is
\begin{equation}
\frac{\deriv n(>\Delta \nu_{\rm obs})}{\deriv z} = \frac{\deriv r}{\deriv z} \int \deriv m \, n_b(m) \, \pi r_{\rm cs}^2(m,\Delta \nu_{\rm obs}),
\label{eq:forest-htwo}
\end{equation}
where $r_{\rm cs}$ is the maximum impact parameter for a bubble of mass $m$ to have a spectral length (in observer's units) of $\Delta \nu_{\rm obs}$ and $\deriv r/\deriv z$ is the comoving line element.  For reference, $\Delta r \approx 1.7 (\Delta \nu_{\rm obs}/100 \kHz) \Mpc$ at $z=10$.  

Figure~\ref{fig:htwo} shows the resulting distribution (transformed into differential form) for a variety of ionized fractions at $z=10$ (we have adjusted $\zeta$ in each case so as to recover the desired $\bxion$).  Except at the smallest $\bxion$, the distribution is fairly flat until it cuts off sharply at relatively large widths.    This is because $n_b(m)$ has a well-defined characteristic size $R_b$, corresponding to the cutoff.  Narrower transmission features arise from a mixture of smaller bubbles and lines of sight that pass near the edges of bubbles with $R \approx R_b$.   Because $R_b$ rapidly grows from a few tenths to a few tens of Mpc \citep{furl04-bub}, the transmission gaps are relatively large spectral features. This contrasts with a model in which galaxy \htwo regions remain isolated, which would have gaps at most a few tens of kHz wide:  thus galaxy clustering is extremely important for the distribution of these bubble features (and hence the 21 cm forest provides a sharp test).  Another key feature is that the total number density of distinct gaps remains roughly constant with $\bxion$ (at least until $\bxion \ga 0.75$).  This is because most new ionizations occur through the growth of existing bubbles (which also merge with their neighbors) rather than through the formation of new bubbles \citep{furl05-rec}.  As a result, the most obvious bubble fluctuations will actually occur relatively early in reionization, when bubbles do not carpet the spectrum.  

%%%%%%%%%%%% FIGURE 4: Ionized regions, bias
\begin{figure}
\begin{center}
\resizebox{8cm}{!}{\includegraphics{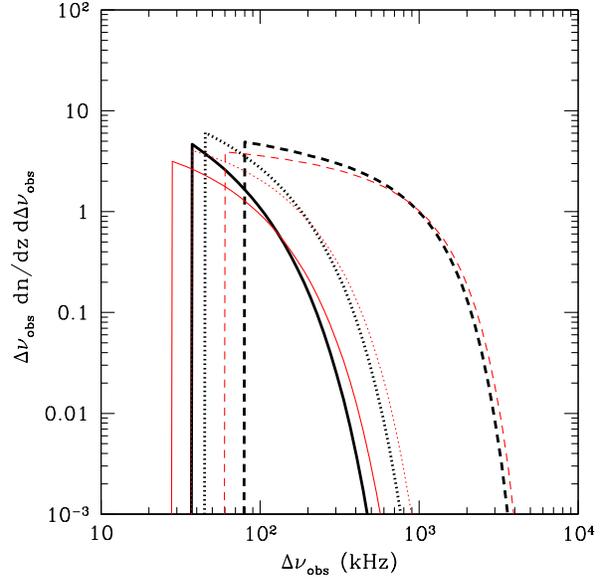}}
\end{center}
%\plotone{f4.eps}
\caption{Differential abundance of \htwo regions intersected per unit redshift and per unit radial width at $z=15$.  The solid, dotted, and dashed lines take $\bxion=0.05,\,0.1,$ and $0.5$, respectively.  The thick and thin curves assume $\zeta \propto m_h^0$ and $m_h^{2/3}$.}
\label{fig:htwo-wgt}
\end{figure}

Of course, the amplitude of the transmission feature is determined by the absorbing neutral gas surrounding the \htwo region (e.g., Figure~\ref{fig:taumean}\emph{c}).  For our fiducial Pop II stars, it drops from $\tau \sim 10^{-3}$ to $\sim 10^{-4}$ over the range $\bxion \sim 0.1$--$0.4$.  Clearly, the contrast will be small even in the early phases of reionization.  Thus, by equation (\ref{eq:forest-sens}), observing \htwo regions would require a source of comparable luminosity to Cygnus A.  On the other hand, if massive Pop III stars dominate, the mean optical depth could be an order of magnitude larger (see Figure~\ref{fig:taumean}\emph{c}).  In this case, Mpc-scale \htwo regions could be observed against mJy radio sources, which may be reasonably common at $z \la 12$ (see \S \ref{disc}).

Figure~\ref{fig:htwo-wgt} shows how the size distribution changes with the parameters of the ionizing sources. The thick curves show an identical model to Figure~\ref{fig:htwo}, except at $z=15$.  The thin curves assume that $\zeta \propto m_{h}^{2/3}$:  thus massive galaxies shift the \htwo regions to larger scales \citep{furl05-charsize}.  This manifests itself in the spectrum by moving the radial cutoff to slightly larger frequency widths; the effect is largest at small $\bxion$, where the bubbles will be easiest to observe.  Obviously the range of $\deriv n/\deriv z$ is relatively small in these models, but measurements will still help to constrain more complicated models (including feedback, for example) to test the basic premises of the \citet{furl04-bub} model, and to measure $\bxion(z)$.

Finally, a comparison of the $\bxion=0.1$ and $0.5$ curves in Figures~\ref{fig:htwo} and \ref{fig:htwo-wgt} shows that the size distribution of \htwo regions at a given ionized fraction does not depend strongly on redshift.  The shapes are nearly identical; the normalization changes primarily through the line element $\deriv r/\deriv z$.

\section{The Cosmic Web} \label{web}

A third type of feature is the exact analog of the \lya forest:  the fluctuating absorption from sheets and filaments beginning to condense out of the IGM.  These can appear as narrow $\tau \ga 1\%$ absorption spikes in the spectrum \citep{carilli02}.  Naively, one might expect that these features would strengthen with time as the cosmic web accumulates more mass through ongoing structure formation.  However, as usual we must also consider the broader context of the radiation backgrounds.  These increase $T_S$ everywhere and hence \emph{decrease} the strength of absorption even inside sheets and filaments.  For example, in the simulations examined by \citet{carilli02}, the number density of absorbers with $\tau > 0.02$ actually decreased from $\sim 50$ to $\sim 4$ over the range $z=10$--$8$.  

To study this problem in a more general context, we follow \citet{schaye01} and consider a cloud with characteristic hydrogen density $n_H$ that is in (or near) hydrostatic equilibrium.  In such a cloud, $\deriv P/\deriv r = -G \, \rho \, M(r)/r^2$, where $P \sim c_s^2 \, \rho$ is the gas pressure, $c_s$ is the sound speed, and $M(r)$ is the mass enclosed within a radius $r$.  Hydrostatic equilibrium thus sets a characteristic ``Jeans length" $L_J$ for the cloud through $c_s^2 \, \rho/L_J \sim G \, \rho^2 \, L_J$, which describes the typical comoving length scale over which the cloud achieves a given overdensity,
\begin{equation}
L_J \equiv \frac{c_s}{\sqrt{G \, \rho}} \sim 7.1 \, (1+\delta)^{-1/2} \, \left( \frac{T_K}{100 \kel} \, \frac{10}{1+z} \, \frac{0.15}{\Omega_m h^2} \right)^{1/2} \kpc.
\label{eq:ljeans}
\end{equation}
Despite its obvious approximations and shortcomings, this simple model does a surprisingly good job describing the \lya forest at $z \sim 3$, which is composed of the same sheets and filaments as the 21 cm forest \citep{schaye01}.  We therefore expect it to provide a reasonable description of high-redshift features as well.  

The column density of the cloud is then $N_{\rm HI} \sim n_H \, L_J$, and we can compute its optical depth via equation (\ref{eq:taucloud}).  Assuming pure thermal broadening (additional sources can only decrease the peak $\tau$) and $T_S \approx T_K$ (valid at high densities and/or for a substantial soft-UV background), we find that the mean overdensity corresponding to a feature of optical depth $\tau$ is
\begin{eqnarray}
1 + \delta & \sim & 94 \, \left( \frac{\tau}{0.01} \right)^2 \, \left( \frac{T_S}{100 \kel} \right)^2 \, \left( \frac{10}{1+z} \right)^3 \nonumber \\
& & \times  \left( \frac{\Omega_m h^2}{0.15} \right) \, \left( \frac{\Omega_b h^2}{0.023} \right)^2.
\label{eq:filament-tau}
\end{eqnarray}
Overdensities $\sim 100$ only occur in virialized objects.  From Figs.~\ref{fig:taumean}\emph{a} and \ref{fig:taumean-fb}\emph{a}, we see that in most cases $T_S \ga 100 \kel$ long before reionization enters its rapid phase, even if X-ray heating is relatively weak.  Clearly, in these sorts of models cosmic web absorption features will be extremely rare during reionization.   Feedback only makes the prospects worse by pushing heating even further back in time relative to the end of reionization.

Recall that \citet{carilli02} found significant absorption even at $z \sim 10$, which may appear to disagree with our claims.  This is because their simulation ignored X-ray heating so had $T_S \sim 30 \kel$ throughout much of the IGM at $z=10$.  Our models show that X-rays can easily surpass this threshold by $z \sim 12$--$15$.  On the other hand, they had $T_S > 100 \kel$ everywhere by $z=8$ -- and indeed the number density of absorbers fell by a factor of ten over that time interval.  So our principal conclusion, that $T_S$  must remain small in order for sheets and filaments to be visible, agrees with their simulation.  This just illustrates the crucial importance of considering the signal in the broadest possible context.

\section{Minihalo Absorption} \label{minihalo}

A final set of features come from minihalos.  These gas clouds can have extremely large column densities and hence produce relatively strong absorption.  \citet{furl02-21cm} computed the absorption profiles of minihalos and their abundance for a particular thermal history (based on the simulation of \citealt{carilli02}).  As above, our goal here is simply to consider the signal in more general models.  

We assume that minihalos form with a comoving number density $n_h(m)$ given by the \citet{press74} mass function (using the \citealt{sheth99} mass function affects our results by $\la 25\%$).  We assume that the gas settles into an isothermal distribution (at the virial temperature) inside the dark matter potential, which follows the \citet{navarro97} distribution.  Other gas profiles will modify the absorption statistics, though not dramatically.  The excess optical depth along a line of sight with impact parameter $\alpha$ through a minihalo is \citep{furl02-21cm}
\begin{equation}
\tau_{\rm mh}(\nu) = 1.91 \times 10^8 \int_0^{R_{\rm max}} \deriv R \, \frac{n_{\rm HI}(r)}{T_S(r) \, b} \, \exp \left[ \frac{-v^2(\nu)}{b^2} \right],
\label{eq:mhtau}
\end{equation}
where $b^2 \equiv 2 k_B T_{\rm vir}/m_p$ is the Doppler parameter, $v(\nu) = c(\nu-\nu_0)/\nu_{21}$, $\nu$ is the frequency at the minihalo (without the cosmological redshifting to the observer), $\nu_{21}=1420 \MHz$, $r^2 = \alpha^2 + R^2$, $R_{\rm max}^2 = r_{\rm vir}^2 - \alpha^2$, and $R$ is measured in physical kiloparsecs.   Here we have neglected the infall region around each minihalo, which can add a significant amount of absorption \citep{furl02-21cm} but is difficult to model accurately.  Note that, because the density varies with radius, $T_S=T_S(r)$ even though the gas is isothermal.  The peak optical depth can be large -- tens of percent -- though lines of sight with $\alpha \ga 0.1 r_{\rm vir}$ usually have $\tau \la 0.02$ \citep{furl02-21cm}.  The typical line widths (resulting from thermal broadening) are $\Delta \nu_{\rm obs} \sim 2 \kHz$.  

As with filaments, the UV background will affect $T_S$ inside minihalos.  However, in this case it has only a modest effect because the local densities and temperatures are large, so that $T_S \sim T_K$ is a reasonable approximation everywhere.  Nevertheless, even if $\tilde{x}_\alpha \gg 1$ everywhere, the minihalo optical depth greatly exceeds that of the IGM, because of their high column densities.  In such a situation, the only difference is that equation (\ref{eq:mhtau}) then measures the excess optical depth over the baseline IGM level (at least to zeroth order).  This marks an important contrast with attempts to study minihalo emission \citep{iliev02, iliev03}:  even when IGM emission overwhelms minihalos \citep{furl06-mh}, they are still distinguishable in absorption through their large columns.

The abundance of minihalo features is then
\begin{equation}
\frac{\deriv n(>\tau_{\rm mh})}{\deriv z} = \frac{\deriv r}{\deriv z} \int_{\mmin}^{m_4} \deriv m \, n_h(m) \, \pi \, r_{\tau}^2(m,\tau_{\rm mh}),
\label{eq:mhfreq}
\end{equation}
where $r_{\tau}(m,\tau_{\rm mh})$ is the maximum impact parameter (in comoving units) to have an optical depth greater than $\tau_{\rm mh}$.  The maximum minihalo mass $m_4$ is assumed to have $T_{\rm vir}=10^4 \kel$, where atomic line cooling becomes efficient and stars can form (e.g., \citealt{barkana01}).  The minimum mass is more problematic.  The Jeans mass obviously increases as X-rays (or other processes) heat the IGM (see, e.g., \citealt{gnedin98}).  But the quantitative effects of thermal feedback remain controversial.  \citet{oh03-entropy} argued that X-rays produce an ``entropy floor" that efficiently suppresses accretion onto minihalos, and they showed that 21 cm absorption becomes much weaker in such a scenario.  However, if the energy is injected after the gas has already achieved relatively high densities, X-ray heating may actually have only a modest effect on minihalo formation.  Perhaps the most interesting implication of minihalo measurements will be to constrain this feedback, so we will consider a range of efficiencies.  For the sake of transparency, we simply assign the minimum mass by the Jeans mass corresponding to a particular IGM temperature.

\begin{figure}
\begin{center}
\resizebox{8cm}{!}{\includegraphics{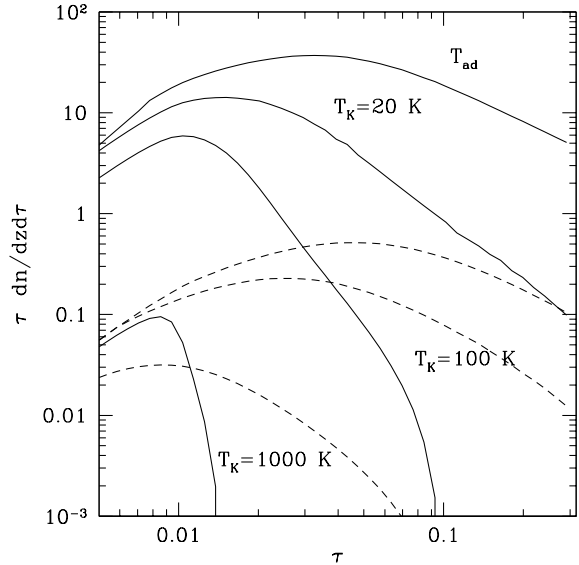}}
\end{center}
%\plotone{f5.eps}
\caption{Differential number density of minihalo absorption features at $z=10$ (solid curves) and $z=20$ (dashed curves).  From top to bottom, each set takes the minimum minihalo mass to be that of a uniform medium at $T_K=T_{\rm ad}(z),\,20,\, 100$ and $1000 \kel$.  Here $T_{\rm ad}(z)$ is the temperature without including any heating from structure formation.  Note that, at $z=20$, the $T_K=10^3 \kel$ curve lies below the plot range.}
\label{fig:forest-mh}
\end{figure}

Figure~\ref{fig:forest-mh} shows the number density of minihalo absorbers at $z=10$ and $20$ (solid and dashed lines, respectively).  At each redshift, the top curves assume $T_K=T_{\rm ad}(z)$ (the temperature in the absence of structure formation processes, computed from RECFAST; \citealt{seager99}).  The remaining curves take $T_K=20,\,100,$ and $1000 \kel$, from top to bottom within each set.  In a cold IGM, the number of minihalo features is large (with $\ga 100$ absorbers per unit redshift in this $\tau$ range at $z=10$) and they can be rather strong (with $\ga 20$ $\tau>0.1$ absorbers per unit redshift). There are about an order of magnitude fewer features at $z=20$, simply because structure formation is so much less advanced at that time.

%%%%%%%%%%%% FIGURE 5: Minihalos

But the number of absorbers decreases rapidly as $T_K$ increases, especially for large optical depths.  This is because the smallest minihalos produce the strongest features: by equation (\ref{eq:taucloud}), $\tau_{\rm mh} \propto \Delta_{\rm vir} r_{\rm vir}/T_S^{3/2} \propto T_{\rm vir}^{-1}$, where $\Delta_{\rm vir}$ is the mean overdensity in virialized objects \citep{barkana01}.   Thus raising the IGM temperature by even a small amount eliminates the strongest features.  On the other hand, massive minihalos produce the more common $\tau \sim 0.01$ absorption features, so they remain even if thermal feedback is strong.  The sensitivity of the (strong) minihalo features tells us that 21 cm absorption can indeed be used to quantify the efficiency of thermal feedback and thus to constrain models of early structure formation.  If these objects continue to form throughout the reionization era, they may therefore present the best possibility for strong features in the 21 cm forest.

\section{Discussion} \label{disc}

We have calculated four kinds of ``21 cm forest" absorption signatures.  First, we computed the evolution of the mean optical depth in a set of plausible histories of early structure formation.  We found that the optical depth can be relatively large ($\tau \ga 0.03$) at sufficiently early epochs, but it begins to decline rapidly once X-ray heating and \lya coupling both become efficient (because $\tau \propto T_S^{-1}$).  By the time reionization begins in earnest, $\tau \sim 10^{-3}$ if Pop II stars dominate the radiation background.  The optical depth can be somewhat higher if very massive Pop III stars dominate, because they cause weaker \lya coupling.

Thus the best bet for observing the 21 cm forest is well before reionization.  \htwo regions may be the easiest features to observe, because they are relatively large ($\ga 10$--$100 \kHz$) even in these early stages (although the contrast with neutral gas may still be relatively small).  Their detection would allow us to measure the size distribution of ionized bubbles, a key quantity in understanding reionization. Sheets and filaments in the cosmic web, as well as minihalos, produce strong but narrow ($\la 2 \kHz$) features.  The vast majority of cosmic web features most likely fade well before reionization, because only virial overdensities can produce strong absorption once $T_S \ga 100 \kel$.  Minihalos may persist longer, because they have much higher column densities.  We expect that up to $\sim 100 $ absorbers with $\tau \ga 0.01$ may be visible per unit redshift at $z=10$ (or $\sim 8$ at $z=20$), and their abundance provides a sensitive measure of thermal feedback \citep{oh03-entropy}.  They may also provide insight into the small-scale matter power spectrum, if the astrophysics can be separated \citep{iliev02, iliev03}.  Regardless, the 21 cm forest avoids the confusion between minihalos and the IGM that plagues tomography measurements \citep{furl06-mh}.

We have considered each of these absorber classes in isolation, but of course in reality they will be intermingled.  Sheets, filaments, and minihalos will fluctuate around the baseline absorption provided by the mean signal, while \htwo regions will appear as gaps (although minihalos may still appear in such regions until they are photoevaporated; \citealt{shapiro04, iliev05}).  This will actually make the ionized bubbles easier to detect, because the gaps will be more obvious even if the mean level of absorption is small (i.e., both the mean level and variance of absorption will decrease in \htwo regions).  On the other hand, it may be somewhat difficult to separate minihalo and cosmic web absorbers, because both produce narrow, strong lines.  

The 21 cm forest is a much simpler observation than large-scale 21 cm tomography, because we can essentially ignore all of the wide-field imaging and foreground problems that plague those observations.  Thus the feasibility of observing the 21 cm forest boils down to two issues:  raw sensitivity and the existence of background sources.  Equation (\ref{eq:forest-sens}) shows that the required source brightness is rather large.  As a result, large collecting areas -- comparable to the Square Kilometer Array -- are required for meaningful constraints.  Somewhat fainter objects can be used to detect fluctuations statistically (for example, minihalos and the cosmic web provide an extra source of apparent noise); however, such games will only buy a factor of few in source flux, so the limits are still relatively stringent \citep{carilli02}.

The most promising candidates for such bright objects ($S \ga 6 \mJy$) are radio-loud quasars; gamma-ray bursts and hypernovae most likely do not reach mJy fluxes \citep{ioka05}.  Do bright radio-loud quasars exist at high redshifts?  Assuming that their number density traces that of optical quasars, the sky could contain $\sim 1000$ $(10)$ sources at $z>8$ $(12)$ \citep{carilli02}.  More detailed models, calibrated to the lower-redshift radio luminosity function, predict $\sim 2000$ sources across the sky with $S>6 \mJy$ at $8<z<12$ \citep{haiman04}, although this prediction is quite sensitive to the assumed radio-loudness parameter.  Thus it appears quite reasonable to expect a fairly large number of appropriate lines of sight with $z <12$.  Intriguingly, such bright sources could already have been detected in higher frequency surveys, such as FIRST \citep{becker95}.  The difficulty lies in separating them from the much more numerous low-redshift interlopers (e.g., \citealt{haiman04}).

Unfortunately, at extremely high redshifts ($z \ga 12$) we will likely be limited to just a few lines of sight.  This is particularly unfortunate because our models show that high optical-depth regions most likely only exist at such epochs.  X-ray heating and Wouthuysen-Field coupling generally precede reionization by a long time interval \citep{furl06-glob}, decreasing the optical depth during reionization.  Thus the rapidly declining source density and the increasing intrinsic signal work against each other, which may ultimately limit the usefulness of the 21 cm forest.  Of course, this is not coincidental:  bright quasars are most likely associated with massive galaxies, which in turn must be rare until reionization gets underway.  Nevertheless, even a few high-redshift lines of sight would be true prizes, because they would allow us to observe the first phases of structure formation with unprecedented spatial resolution and to constrain the radiation backgrounds in the early Universe.  In this context, the recent downward revision of the \emph{WMAP} optical depth \citep{spergel06, page06} is certainly good news, because it suggests that heating and spin temperature coupling may have occurred at somewhat later times, and the requirements on background sources may be somewhat less stringent.  Given the many uncertainties about the $z \ga 6$ Universe, lower-redshift sources will also offer meaningful constraints (for example, ruling out weak spin-temperature coupling from massive Pop III stars and testing feedback models for minihalos), even if the most plausible cases  have considerably smaller signals.

I thank the Tapir group at Caltech for hospitality while some of this work was completed.

%\bibliographystyle{apj}
%\bibliography{Ref_21cm}

\begin{thebibliography}{58}
\expandafter\ifx\csname natexlab\endcsname\relax\def\natexlab#1{#1}\fi

\bibitem[{{Barkana} \& {Loeb}(2001)}]{barkana01}
{Barkana}, R., \& {Loeb}, A. 2001, \physrep, 349, 125

\bibitem[{{Barkana} \& {Loeb}(2005)}]{barkana05-ts}
---. 2005, \apj, 626, 1

\bibitem[{{Becker} {et~al.}(1995){Becker}, {White}, \& {Helfand}}]{becker95}
{Becker}, R.~H., {White}, R.~L., \& {Helfand}, D.~J. 1995, \apj, 450, 559

\bibitem[{{Bouwens} {et~al.}(2004)}]{bouwens04}
{Bouwens}, R.~J., {et~al.} 2004, \apjl, 616, L79

\bibitem[{{Carilli} {et~al.}(2002){Carilli}, {Gnedin}, \& {Owen}}]{carilli02}
{Carilli}, C.~L., {Gnedin}, N.~Y., \& {Owen}, F. 2002, \apj, 577, 22

\bibitem[{{Chen} \& {Miralda-Escud{\' e}}(2004)}]{chen04}
{Chen}, X., \& {Miralda-Escud{\' e}}, J. 2004, \apj, 602, 1

\bibitem[{{Ciardi} \& {Ferrara}(2005)}]{ciardi05-rev}
{Ciardi}, B., \& {Ferrara}, A. 2005, Space Science Reviews, 116, 625

\bibitem[{{Ciardi} {et~al.}(2005){Ciardi}, {Scannapieco}, {Stoehr}, {Ferrara},
  {Iliev}, \& {Shapiro}}]{ciardi05-mh}
{Ciardi}, B., {Scannapieco}, E., {Stoehr}, F., {Ferrara}, A., {Iliev}, I.~T.,
  \& {Shapiro}, P.~R. 2005, \mnras, in press (astro-ph/0511623)

\bibitem[{{Dijkstra} {et~al.}(2004){Dijkstra}, {Haiman}, {Rees}, \&
  {Weinberg}}]{dijkstra04-feed}
{Dijkstra}, M., {Haiman}, Z., {Rees}, M.~J., \& {Weinberg}, D.~H. 2004, \apj,
  601, 666

\bibitem[{{Fan} {et~al.}(2001)}]{fan01}
{Fan}, X., {et~al.} 2001, \aj, 122, 2833

\bibitem[{{Fan} {et~al.}(2005)}]{fan06}
---. 2005, in press at \aj \ (astro-ph/0512080)

\bibitem[{{Field}(1958)}]{field58}
{Field}, G.~B. 1958, Proc. I.R.E., 46, 240

\bibitem[{{Field}(1959{\natexlab{a}})}]{field59-obs}
---. 1959{\natexlab{a}}, \apj, 129, 525

\bibitem[{{Field}(1959{\natexlab{b}})}]{field59-ts}
---. 1959{\natexlab{b}}, \apj, 129, 536

\bibitem[{{Furlanetto}(2006)}]{furl06-glob}
{Furlanetto}, S.~R. 2006, submitted to \mnras \ (astro-ph/0604040)

\bibitem[{{Furlanetto} \& {Loeb}(2002)}]{furl02-21cm}
{Furlanetto}, S.~R., \& {Loeb}, A. 2002, \apj, 579, 1

\bibitem[{{Furlanetto} \& {Loeb}(2004)}]{furl04-sh}
---. 2004, \apj, 611, 642

\bibitem[{{Furlanetto} \& {Loeb}(2005)}]{furl05-double}
---. 2005, \apj, 634, 1

\bibitem[{{Furlanetto} {et~al.}(2006{\natexlab{a}}){Furlanetto}, {McQuinn}, \&
  {Hernquist}}]{furl05-charsize}
{Furlanetto}, S.~R., {McQuinn}, M., \& {Hernquist}, L. 2006{\natexlab{a}},
  \mnras, 365, 115

\bibitem[{{Furlanetto} \& {Oh}(2005)}]{furl05-rec}
{Furlanetto}, S.~R., \& {Oh}, S.~P. 2005, \mnras, 363, 1031

\bibitem[{{Furlanetto} \& {Oh}(2006)}]{furl06-mh}
---. 2006, submitted to \apj \ 

\bibitem[{{Furlanetto} {et~al.}(2006{\natexlab{b}}){Furlanetto}, {Oh}, \&
  {Briggs}}]{furl06-review}
{Furlanetto}, S.~R., {Oh}, S.~P., \& {Briggs}, F.~H. 2006{\natexlab{b}},
  Physics Reports, in preparation

\bibitem[{{Furlanetto} {et~al.}(2004){Furlanetto}, {Zaldarriaga}, \&
  {Hernquist}}]{furl04-bub}
{Furlanetto}, S.~R., {Zaldarriaga}, M., \& {Hernquist}, L. 2004, \apj, 613, 1

\bibitem[{{Gilfanov} {et~al.}(2004){Gilfanov}, {Grimm}, \&
  {Sunyaev}}]{gilfanov04}
{Gilfanov}, M., {Grimm}, H.-J., \& {Sunyaev}, R. 2004, \mnras, 347, L57

\bibitem[{{Gnedin} \& {Hui}(1998)}]{gnedin98}
{Gnedin}, N.~Y., \& {Hui}, L. 1998, \mnras, 296, 44

\bibitem[{{Grimm} {et~al.}(2003){Grimm}, {Gilfanov}, \& {Sunyaev}}]{grimm03}
{Grimm}, H.-J., {Gilfanov}, M., \& {Sunyaev}, R. 2003, \mnras, 339, 793

\bibitem[{{Haiman} {et~al.}(2004){Haiman}, {Quataert}, \& {Bower}}]{haiman04}
{Haiman}, Z., {Quataert}, E., \& {Bower}, G.~C. 2004, \apj, 612, 698

\bibitem[{{Hirata}(2005)}]{hirata05}
{Hirata}, C.~M. 2005, submitted to \mnras \ (astro-ph/0507102)

\bibitem[{{Hogan} \& {Rees}(1979)}]{hogan79}
{Hogan}, C.~J., \& {Rees}, M.~J. 1979, \mnras, 188, 791

\bibitem[{{Iliev} {et~al.}(2005{\natexlab{a}}){Iliev}, {Mellema}, {Pen},
  {Merz}, {Shapiro}, \& {Alvarez}}]{iliev05-sim}
{Iliev}, I.~T., {Mellema}, G., {Pen}, U.-L., {Merz}, H., {Shapiro}, P.~R., \&
  {Alvarez}, M.~A. 2005{\natexlab{a}}, submitted to \mnras (astro-ph/0512187)

\bibitem[{{Iliev} {et~al.}(2003){Iliev}, {Scannapieco}, {Martel}, \&
  {Shapiro}}]{iliev03}
{Iliev}, I.~T., {Scannapieco}, E., {Martel}, H., \& {Shapiro}, P.~R. 2003,
  \mnras, 341, 81

\bibitem[{{Iliev} {et~al.}(2005{\natexlab{b}}){Iliev}, {Scannapieco}, \&
  {Shapiro}}]{iliev05}
{Iliev}, I.~T., {Scannapieco}, E., \& {Shapiro}, P.~R. 2005{\natexlab{b}},
  \apj, 624, 491

\bibitem[{{Iliev} {et~al.}(2002){Iliev}, {Shapiro}, {Ferrara}, \&
  {Martel}}]{iliev02}
{Iliev}, I.~T., {Shapiro}, P.~R., {Ferrara}, A., \& {Martel}, H. 2002, \apjl,
  572, L123

\bibitem[{{Ioka} \& {M{\' e}sz{\' a}ros}(2005)}]{ioka05}
{Ioka}, K., \& {M{\' e}sz{\' a}ros}, P. 2005, \apj, 619, 684

\bibitem[{{Kitayama} \& {Ikeuchi}(2000)}]{kitayama00}
{Kitayama}, T., \& {Ikeuchi}, S. 2000, \apj, 529, 615

\bibitem[{{Kuhlen} {et~al.}(2006){Kuhlen}, {Madau}, \&
  {Montgomery}}]{kuhlen06-21cm}
{Kuhlen}, M., {Madau}, P., \& {Montgomery}, R. 2006, \apjl, 637, L1

\bibitem[{{Liszt}(2001)}]{liszt01}
{Liszt}, H. 2001, \aap, 371, 698

\bibitem[{{Madau} {et~al.}(1997){Madau}, {Meiksin}, \& {Rees}}]{madau97}
{Madau}, P., {Meiksin}, A., \& {Rees}, M.~J. 1997, \apj, 475, 429

\bibitem[{{Malhotra} \& {Rhoads}(2004)}]{malhotra04}
{Malhotra}, S., \& {Rhoads}, J.~E. 2004, \apjl, 617, L5

\bibitem[{{Navarro} {et~al.}(1997){Navarro}, {Frenk}, \& {White}}]{navarro97}
{Navarro}, J.~F., {Frenk}, C.~S., \& {White}, S.~D.~M. 1997, \apj, 490, 493
  (NFW)

\bibitem[{{Oh} \& {Haiman}(2003)}]{oh03-entropy}
{Oh}, S.~P., \& {Haiman}, Z. 2003, \mnras, 346, 456

\bibitem[{{Page} {et~al.}(2006)}]{page06}
{Page}, L., {et~al.} 2006, submitted to \apj \ (astro-ph/0603450)

\bibitem[{{Press} \& {Schechter}(1974)}]{press74}
{Press}, W.~H., \& {Schechter}, P. 1974, \apj, 187, 425

\bibitem[{{Pritchard} \& {Furlanetto}(2005)}]{pritchard05}
{Pritchard}, J.~R., \& {Furlanetto}, S.~R. 2005,  \mnras, in press 
  (astro-ph/0508381)

\bibitem[{{Ranalli} {et~al.}(2003){Ranalli}, {Comastri}, \&
  {Setti}}]{ranalli03}
{Ranalli}, P., {Comastri}, A., \& {Setti}, G. 2003, \aap, 399, 39

\bibitem[{{Schaye}(2001)}]{schaye01}
{Schaye}, J. 2001, \apj, 559, 507

\bibitem[{{Scott} \& {Rees}(1990)}]{scott90}
{Scott}, D., \& {Rees}, M.~J. 1990, \mnras, 247, 510

\bibitem[{{Seager} {et~al.}(1999){Seager}, {Sasselov}, \& {Scott}}]{seager99}
{Seager}, S., {Sasselov}, D.~D., \& {Scott}, D. 1999, \apjl, 523, L1

\bibitem[{{Shapiro} {et~al.}(2004){Shapiro}, {Iliev}, \& {Raga}}]{shapiro04}
{Shapiro}, P.~R., {Iliev}, I.~T., \& {Raga}, A.~C. 2004, \mnras, 348, 753

\bibitem[{{Sheth} \& {Tormen}(1999)}]{sheth99}
{Sheth}, R.~K., \& {Tormen}, G. 1999, \mnras, 308, 119

\bibitem[{{Smith}(1966)}]{smith66}
{Smith}, F.~J. 1966, Plan. Space Sci., 14, 929

\bibitem[{{Spergel} {et~al.}(2006)}]{spergel06}
{Spergel}, D.~N., {et~al.} 2006, submitted to \apj \ (astro-ph/0603449)

\bibitem[{{Taniguchi} {et~al.}(2005)}]{taniguchi05}
{Taniguchi}, Y., {et~al.} 2005, \pasj, 57, 165

\bibitem[{{Thoul} \& {Weinberg}(1996)}]{thoul96}
{Thoul}, A.~A., \& {Weinberg}, D.~H. 1996, \apj, 465, 608

\bibitem[{{Wouthuysen}(1952)}]{wouthuysen52}
{Wouthuysen}, S.~A. 1952, \aj, 57, 31

\bibitem[{{Zahn} {et~al.}(2006)}]{zahn06-comp}
{Zahn}, O., {et~al.} 2006, in preparation

\bibitem[{{Zaldarriaga} {et~al.}(2004){Zaldarriaga}, {Furlanetto}, \&
  {Hernquist}}]{zald04}
{Zaldarriaga}, M., {Furlanetto}, S.~R., \& {Hernquist}, L. 2004, \apj, 608, 622

\bibitem[{{Zygelman}(2005)}]{zygelman05}
{Zygelman}, B. 2005, \apj, 622, 1356

\end{thebibliography}

\end{document}